# Magnetic and electron transport properties of the rare-earth cobaltates, La$_{0.7-x}$Ln$_x$Ca$_{0.3}$CoO$_3$ (Ln = Pr, Nd, Gd and Dy) : A case of phase separation

Asish K Kundu[a], K Ramesha[b], Ram Seshadri[b] and C N R Rao[a]*

[a]*Chemistry and Physics of Materials Unit, Jawaharlal Nehru Centre for Advanced Scientific Research, Jakkur P.O., Bangalore - 560064, India.*
[b]*Materials Department, University of California, Santa Barbara, CA – 93106, USA.*

**Abstract**

Magnetic and electrical properties of four series of rare earth cobaltates of the formula La$_{0.7-x}$Ln$_x$Ca$_{0.3}$CoO$_3$ with Ln = Pr, Nd, Gd and Dy have been investigated. Compositions close to x = 0.0 contain large ferromagnetic clusters or domains, and show Brillouin-like behaviour of the field-cooled DC magnetization data with fairly high ferromagnetic T$_c$ values, besides low electrical resistivities with near-zero temperature coefficients. The zero-field-cooled data generally show a non-monotonic behaviour with a peak at a temperatures slightly lower than T$_c$. The near x = 0.0 compositions show a prominent peak corresponding to the T$_c$ in the AC-susceptibility data. The ferromagnetic T$_c$ varies linearly with x or the average radius of the A-site cations, <r$_A$>. With increase in x or decrease in <r$_A$>, the magnetization value at any given temperature decreases markedly and the AC-susceptibility measurements show a prominent transition arising from small magnetic clusters with some characteristics of a spin-glass. Electrical resistivity increases with increase in x, showed a significant increase around a critical value of x or <r$_A$>, at which composition the small clusters also begin to dominate. These properties can be understood in terms of a phase separation scenario wherein large magnetic clusters give way to smaller ones with increase in x, with both types of clusters being present in certain compositions. The changes in magnetic and electrical properties occur parallely since the large ferromagnetic clusters are hole-rich and the small clusters are hole-poor. Variable-range hopping seems to occur at low temperatures in these cobaltates.

* For correspondence: cnrrao@jncasr.ac.in

# 1. Introduction

Rare-earth manganates of the general formula $Ln_{1-x}A_xMnO_3$ (Ln = rare earth, A = alkaline earth), possessing the perovskite structure, exhibit interesting properties such as colossal magnetoresistance, charge ordering and electronic phase separation [1-5]. These properties of the manganates are strongly influenced by the average radius of the A-site cations, $<r_A>$. It has been found recently that manganates of the type $La_{0.7-x}Ln_xCa_{0.3}MnO_3$ (Ln = Pr, Nd, Gd, Y) undergo electronic phase separation beyond a critical composition $x_c$. While the $x < x_c$ compositions exhibit ferromagnetism and metallicity, materials with $x > x_c$ are non-magnetic insulators, with the magnetic moment dropping sharply in these compositions [6 - 9]. Such changes in properties in a series of manganates where the carrier concentration or the $Mn^{4+}/Mn^{3+}$ ratio is kept constant is indeed noteworthy. Perovskite cobaltates of the type $Ln_{1-x}A_xCoO_3$ (Ln = rare earth, A = alkaline earth) are in many ways similar to the manganates [10 - 13]. Accordingly, many of the cobaltates show ferromagnetism and metallicity depending on the composition and the size of the A-site cations. The cobaltates also show some unusual features in their magnetic properties. Thus Itoh et al. [14] find that $La_{1-x}Sr_xCoO_3$ shows spin-glass behaviour when $0.0 \leq x \leq 0.18$ and cluster-glass when $0.18 \leq x \leq 0.5$. Ganguly et al. [15] reported long-range freezing of superparamagnetic clusters when $x < 0.3$. For $x = 0.5$, a cluster-glass magnetic behaviour has been reported by Kumar et al. [16] who ascribe this property to magnetocrystalline anisotropy. Wu and Leighton [17] have recently examined the properties of $La_{1-x}Sr_xCoO_3$ and suggest that the system is best described as one dominated by glassy ferromagnetism and magnetic phase separation. We considered it important to investigate the magnetic and electron transport properties of cobaltates of the type $La_{0.7-x}Ln_xCa_{0.3}CoO_3$ (Ln = Pr, Nd, Gd or Dy), in view of the comparable nature of the cobaltates and the manganates, and also because of the likelihood that phase separation may occur in the cobaltate system as well because of the presence of strongly correlated electrons [5]. While the carrier concentration remains constant in $La_{0.7-x}Ln_xCa_{0.3}CoO_3$, the $<r_A>$ and the associated factors vary. Furthermore, the Ca-substituted cobaltates were considered more likely to exhibit phase separation and related effects due to the smaller $<r_A>$ or the $e_g$ band-width compared to the Sr-substituted materials.

**2. Experimental procedure**

Polycrystalline samples of $La_{0.7-x}Ln_xCa_{0.3}CoO_3$ (Ln = Pr, Nd, Gd, or Dy), were prepared by the conventional ceramic method. Stoichiometric mixtures of the respective rare earth oxides, alkaline earth carbonates and $Co_3O_4$ were weighed in desired proportions and milled for few hours with propanol. After the mixed powders were dried, they were calcined in air at 950°C followed by heating at 1000 and 1100°C for 12 h in air. The powders thus obtained were pelletized and the pellets sintered at 1200°C for 24 h in an oxygen atmosphere. To improve the oxygen stoichiometry, the samples were annealed in an oxygen atmosphere below 900°C and cooled gradually. The oxygen stoichiometry was determined by iodometric titrations. The error in the determination of oxygen content was ± 0.02. The oxygen stoichiometry in the cobaltates studied by us was 2.97 ± 0.03.

The phase purity of the samples was established by recording the X-ray diffraction patterns in the 2θ range of 10°- 80° with a Seiferts 3000 TT diffractometer, employing Cu-Kα radiation. The unit cell parameters of $La_{0.7-x}Ln_xCa_{0.3}CoO_3$ (Ln = Pr,

Nd, Gd, Dy) are listed in table 1 along with the weighted average radius $\langle r_A \rangle$. The $\langle r_A \rangle$ values were calculated using the Shannon radii for 12-coordination in the case of rhombohedral cobaltates, and for 9-coordination in the case of the orthorhombic ones. Magnetization (M) measurements were made over the 5 – 300 K range with a vibrating sample magnetometer (Lakeshore 7300) and with a Quantum Design MPMS 5XL magnetometer. Electrical resistivity ($\rho$) measurements were carried out in the 20 - 300 K range by the four-probe method with silver epoxy electrodes.

## 3. Results and discussion

In figure 1, we show the temperature dependence of the DC magnetization of a few compositions of $La_{0.7-x}Ln_xCa_{0.3}CoO_3$ (Ln = Pr, Nd). Both the zero-field-cooled (ZFC) and field-cooled (FC) data are shown in the figure. The FC curve of the x = 0.0 composition shows a Brillouin-type temperature dependence, and a $T_c$ value of around 170 K. With increase in x, the ZFC data show lower magnetization values and a non-monotonic behaviour. Accordingly, we see a peak in the ZFC data, at a temperature, $T_P$, the peak temperature being slightly lower than the $T_c$ value from the FC data. The maximum value of the magnetization attained as well as the apparent $T_c$ value decrease with increasing x or decreasing $\langle r_A \rangle$. At large x, we notice the appearance of a shoulder-like feature in the magnetization curve, the feature becoming clearly noticeable at x ≥ 0.6 and x ≥ 0.4 respectively in the Pr- and Nd-substituted series of cobaltates. This feature can be seen clearly in figure 2 where the data of $Pr_{0.7}Ca_{0.3}CoO_3$ (x = 0.7) are presented. Accordingly, AC-susceptibility measurements of $La_{0.7}Ca_{0.3}CoO_3$ (x = 0.0, $\langle r_A \rangle$ = 1.354 Å) show one major frequency-independent transition around 150 K corresponding to the $T_c$, while $Pr_{0.7}Ca_{0.3}CoO_3$ (x = 0.7, $\langle r_A \rangle$ = 1.179 Å) shows two distinct transitions around 65 K and 35 K, the latter with a greater frequency dependence as shown in figure 3. $La_{0.4}Pr_{0.3}Ca_{0.3}CoO_3$ ($<r_A>$ = 1.194 Å) also shows a single transition corresponding to the $T_c$ which is frequency independent just as $La_{0.7}Ca_{0.3}CoO_3$. The frequency-independent high-temperature transition in $Pr_{0.7}Ca_{0.3}CoO_3$ is due to the large magnetic clusters (the so-called cluster-glass [11, 14]) as in the x = 0.0 composition and the low-temperature transition is due to small magnetic clusters which seem to show some spin-glass characteristics. Thus, with the increase in x or decrease in $<r_A>$ in the $La_{0.7-x}Ln_xCa_{0.3}CoO_3$ system, the large ferromagnetic clusters seem to progressively give way to the small clusters, giving rise to magnetic phase separation. The presence of very weak features at low temperatures in the AC-susceptibility data of $La_{0.7}Ca_{0.3}CoO_3$ (figure 3a) indicates that the proportion of small clusters is negligible at x ≈ 0.0. In $Nd_{0.7}Ca_{0.3}CoO_3$ ($<r_A>$ = 1.168 Å), on the other hand, we only see a frequency-dependent low-temperature transition around 20 K due to the small magnetic clusters (figure 3d).

The magnetic properties of the Gd- and Dy-substituted compositions, $La_{0.7-x}Ln_xCa_{0.3}CoO_3$ (Ln = Gd, Dy) shown in figure 4 are similar to those of the Pr- and Nd-substituted compositions, except that at low temperatures (< 30 K) an increase in magnetization is noticed. Such an increase arising from the contribution from the Gd and Dy ions has been observed earlier [18]. The $T_c$ values obtained from FC data in the four series of cobaltates are plotted against x in figure 5. The $T_c$ value decreases linearly with increasing x. These data can be rationalized in terms of the average radius of A-site cations, $<r_A>$. Thus, a plot of $T_c$ versus $<r_A>$ is sensibly linear as can be seen from the inset of figure 5.

We have recorded M-H curves of the four series of cobaltates and show typical data for two series of compounds in figure 6. The hysteresis curves do not show saturation in all the compositions. The absence of saturation is a characteristic of a glassy system. Furthermore, the remanent magnetization, $M_r$, decreases with increase in x or decrease in $<r_A>$. We have plotted $M_r$ at 50 K against $<r_A>$ in figure 7, where we have also shown the variation of the magnetization, M, with $<r_A>$ at this temperature. Both M and $M_r$ increase with $<r_A>$, but their values become rather low when $<r_A> \lesssim 1.18$ Å.

The electrical resistivities of the cobaltates show trends which are consistent with the magnetic properties. In figure 8 we show typical resistivity data for two series of cobaltates. The temperature coefficient of resistivity changes from a near-zero value to a positive value around $x_c$ in some of the series, but in all the four series the magnitude of resistivity shows a marked increase around a critical composition $x_c$ or a critical radius $<r_A^c>$ of ~1.18 Å. We observe this behaviour at x = 0.49 and 0.195 for the Nd and Dy series respectively (figure 8). In order to rationalize the resistivity data in the four series of cobaltates, we have plotted the resistivity data at 50 K against $<r_A>$ in figure 7c. There is a noticeable increase in the resistivity with decreasing $<r_A>$, with a change in slope around 1.18 Å. It is to be noted that below this value of the A-site cation radius, electronic phase separation and charge-ordering occur in the rare earth manganates [9]. The $<r_A>$ value of 1.18 Å in the cobaltates corresponds to x ≈ 0.6, 0.49, 0.24 and 0.195 respectively in the Pr, Nd, Gd and Dy substituted series of cobaltates and we denote these compositions as $x_c$. It appears that the small magnetic cluster regime becomes prominent around $x_c$ or $<r_A^c>$. Even though the low-temperature peak in the AC-susceptibility data is frequency-dependent, the small magnetic cluster regime at small $<r_A>$ in the La$_{0.7-x}$Ln$_x$Ca$_{0.3}$CoO$_3$ system cannot be strictly considered to be that of a spin-glass since we do not observe a logarithmic or exponential decay of the isothermal remanent magnetization. We can certainly consider the regime to be magnetically inhomogeneous. The inhomogeneous nature of the cobaltates prevails over the entire range of compositions (x = 0.0 – 0.7). It seems that around a composition close to $x_c$ or $<r_A^c>$, there is a marked change in the distribution of the magnetic species. Thus, when $x < x_c$ or $<r_A> > <r_A^c>$, relatively large ferromagnetic clusters or domains are present in the system, resulting in large magnetization and $T_c$ values. When $x > x_c$ or $<r_A> < <r_A^c>$, the magnetic clusters become small in size. The ferromagnetic clusters being hole-rich, the electrical resistivity data show changes around the same compositions as the magnetization data, the compositions with $<r_A> > <r_A^c>$ exhibiting lower resistivities and near-zero temperature coefficients of resistivity. While one may treat the change in the nature of magnetic species around $x_c$ or $<r_A^c>$ as a mere change in size distribution, we consider it more appropriate to treat it as a case of phase separation since we observe more than one transition in the AC-susceptibility data and the resistivity changes parallel the changes in the magnetic properties. The phase separated regime here involves the coexistence of large ferromagnetic clusters which are hole-rich and small clusters which are hole-poor.

It is instructive to examine the nature of the spin-states of cobalt in the La$_{0.7-x}$Ln$_x$Ca$_{0.3}$CoO$_3$ system. We can estimate the magnetic moment of the cobalt ion from the inverse magnetic susceptibility data at high temperatures (180 – 300 K). The susceptibility value as well as the slope of the inverse susceptibility-temperature plot give an average magnetic moment value of 4.5 $\mu_B$ per cobalt ion in all the series of cobaltates.

This value suggests that in the 180 – 300 K range, the cobalt ions are in the intermediate-spin (IS) and or high-spin (HS) states. The IS and HS states of $Co^{3+}$ correspond to the electronic configurations $t_{2g}^5 e_g^1$ (S = 1) and $t_{2g}^4 e_g^2$ (S = 2) respectively, and those of $Co^{4+}$ to $t_{2g}^4 e_g^1$ (S = 3/2) and $t_{2g}^3 e_g^2$ (S = 5/2). Investigations of the spin-state transitions in the cobaltates have shown that at high temperatures, the cobalt ions are mostly in the IS or the HS state [13]. At low temperatures, some of the cobalt ions may go to the low-spin (LS) state, corresponding to the $t_{2g}^6$ (S = 0) and $t_{2g}^5$ (S = 1/2) configurations in $Co^{3+}$ and $Co^{4+}$ ions respectively. The ferromagnetic clusters present prominently at $x < x_c$ or $<r_A>$ > $<r_A^c>$ involve cobalt ions in the IS or HS states. The ferromagnetic regime will therefore be hole-rich, the size of the clusters or the domains decreasing with increasing x or decreasing $<r_A>$. We would, therefore, expect a magnetic percolation threshold as well as electrical percolation in the system. We find that a plot of log ρ versus log $|<r_A> - <r_A^c>|$ is linear with a negative slope of around 0.5 at 50 K.

We have explored whether the resistivity behaviour of $La_{0.7-x}Ln_xCa_{0.3}CoO_3$ conforms to activated hopping, defined by log ρ ∝ $(1/T^n)$ where n = 1, 2 or 4. Here, n = 1 corresponds to a simple Arrhenius behaviour. When n = 2, the hopping is referred to as nearest-neighbour hopping (NNH), controlled by Coulombic forces. When n = 4, there would be variable range hopping (VRH) and the hopping dynamics is controlled by the collective excitation of the charge carriers [19, 20]. The resistivity data in the 49 – 200 K region could be fitted to a $T^{-1/2}$ dependence with the standard deviation varying between 0.008 and 0.016. The standard deviation for the $T^{-1/4}$ fits is generally much smaller (0.005 – 0.012). In figure 9, we show typical fits to the $T^{-1/4}$ law. The occurrence of VRH in the cobaltates is consistent with the earlier studies of Rao et al [21, 22] on the rare earth cobaltates.

## 4. Conclusions

The present investigation of the magnetic and electrical properties of $La_{0.7-x}Ln_xCa_{0.3}CoO_3$ (Ln = Pr, Nd, Gd or Dy) can be understood in terms of a phase separation scenario wherein large carrier-rich ferromagnetic clusters and carrier-poor smaller clusters coexist at some compositions. Accordingly, at large x, we observe two prominent magnetic transitions, the one at low temperatures being associated with the small clusters. Since the ferromagnetic clusters prominent at small x are hole-rich, we observe a change in the electrical resistivity behaviour at a critical value $x_c$, where the size distribution of magnetic clusters undergoes significant changes. The critical value of x in the four series of cobaltates corresponds to the critical value of radius, $<r_A^c>$, of 1.18 Å, a value where rare earth manganates of the type $La_{0.7-x}Ln_xCa_{0.3}MnO_3$ (Ln = Nd, Gd or Y) are known to exhibit charge ordering and phase separation prominently [9]. It appears that around $<r_A^c>$ or $x_c$, a significant change occurs in the $e_g$ band-width and the charge carriers become more localized, causing changes in the magnetic and electron transport properties. It is well to recall that the electrical resistivity and ferromagnetism in the cobaltates are linked to the presence of the $Co^{3+}$-O- $Co^{4+}$ species with the appropriate spin states of cobalt ions.

**Acknowledgements**

The authors thank BRNS (DAE), India, for support of this research. AKK wants to thank the University Grants Commission, India, for a fellowship award. The authors acknowledge the MRL facilities supported by the National Science Foundation under Award No. DMROO-80034.

**References**

[1] Rao C N R and Raveau B (eds), 1998 *Colossal Magnetoresistance, Charge Ordering and Related Properties of Manganese Oxides* (World Scientific: Singapore)

[2] Tokura Y (eds) 1999 *Colossal Magnetoresistance Oxides* (London: Gorden and Breach)

[3] Ramirez A P 1997 *J. Phys: Condens. Matter.* **9** 8171

[4] Dagotto E (eds) 2003 Nanoscale Phase Separation and Colossal Magnetoresistance (Spinger: Berlin)

[5] Rao C N R and Vanitha P V 2002 Curr. Opin. Solid State Mater. Sci. **6** 97

[6] Uehara M, Mori S, Chen C H and Cheong S W 1999 *Nature* **399** 560

[7] Lee H J, Kim K H, Kim M W, Noh T W, Kim B G, Koo T Y, Cheong S W, Wang Y J and Wei X 2002 *Phys. Rev.* B **65** 115118

[8] Balagurov A M, Pomjakushin Y Yu, Sheptyakov D V, Aksenov V L, Fischer P, Keller L, Gorbenko O Yu, Kaul A and Babushkina N A 2001 *Phys. Rev.* B **64** 024420

[9] Sudheendra L and Rao C N R 2003 *J. Phys.: Condens. Matter* **15** 3029

[10] Rao C N R, Parkash O, Bahadur D, Ganguly P and Nagabhushana S 1977 *J. Solid State Chem.* **22** 353

[11] Senaris-Rodriguez M A and Goodenough J B 1995 *J. Solid State Chem.* **118** 323 and the references therein.

[12] Saitoh T, Mizokawa T, Fujimori A, Abbate N, Takeda Y and Takano M 1997 *Phys. Rev.* B **56** 1290

[13] Rao C N R, Seikh M M and Chandrabhas N 2004, *Topics in Current Chemistry* **234**, 1.

[14] Itoh M, Natori I, Kubota S and Matoya K 1994, *J. Phys. Soc. Jpn.* **63** 1486

[15] Ganguly P, Anil Kumar P S, Santosh P N and Mulla I S 1994 *J. Phys.: Condens. Matter* **6** 533

[16] Anil Kumar P S, Joy P A and Date S K 1998 *J. Phys.: Condens. Matter* **10** L487

[17] Wu J and Leighton C 2003 *Phys. Rev.* B **67** 174408

[18] Tong W, Hu L, Zhu H, Tan S and Zhang Y 2004 *J. Phys.: Condens. Matter* **16** 103

[19] Mott N F 1990 *Metal-Insulator Transitions* (Taylor & Francis, London).

[20] Shklovskii B I and Efros A L 1984 *Electronic Properties of Doped Semiconductors* (Springer, Berlin).

[21] Rao C N R and Parkash O 1977 *Phil. Mag.* **35** 1111

[22] Rao C N R, Bhide V G and Mott N F 1975 *Phil. Mag.* **32** 1277

**Figure captions**

**Figure 1.** Temperature variation of the magnetization, M, of (a) La$_{0.7-x}$Pr$_x$Ca$_{0.3}$CoO$_3$ and (b) La$_{0.7-x}$Nd$_x$Ca$_{0.3}$CoO$_3$. ZFC data in broken curves and FC data in solid curves (at 1 kOe).

**Figure 2.** Temperature variation of the magnetization, M, of Pr$_{0.7}$Ca$_{0.3}$CoO$_3$ (at 100 Oe).

**Figure 3.** Temperature variation of the AC-susceptibility of (a) La$_{0.7}$Ca$_{0.3}$CoO$_3$ (b) La$_{0.4}$Pr$_{0.3}$Ca$_{0.3}$CoO$_3$ (c) Pr$_{0.7}$Ca$_{0.3}$CoO$_3$ and (d) Nd$_{0.7}$Ca$_{0.3}$CoO$_3$ at two different frequency.

**Figure 4.** Temperature variation of the magnetization, M, of (a) La$_{0.7-x}$Gd$_x$Ca$_{0.3}$CoO$_3$ and (b) La$_{0.7-x}$Dy$_x$Ca$_{0.3}$CoO$_3$. ZFC data in broken curves and FC data in solid curves (1 kOe).

**Figure 5.** Variation of the ferromagnetic T$_c$ with x in La$_{0.7-x}$Ln$_x$Ca$_{0.3}$CoO$_3$. The inset shows the variation of T$_c$ with $\langle r_A \rangle$ (Å).

**Figure 6.** Typical hysteresis curves of La$_{0.7-x}$Ln$_x$Ca$_{0.3}$CoO$_3$ at 5 K (Ln = Pr, Nd).

**Figure 7.** Variation of (a) the magnetic moment, µ$_B$, (b) remanent magnetization, M$_r$, and (c) the electrical resistivity, ρ, in La$_{0.7-x}$Ln$_x$Ca$_{0.3}$CoO$_3$ at 50 K with $\langle r_A \rangle$ (Å) (Ln = Pr, Nd, Gd and Dy).

**Figure 8.** Temperature variation electrical resistivity, ρ, of La$_{0.7-x}$Ln$_x$Ca$_{0.3}$CoO$_3$ (Ln = Nd and Dy).

**Figure 9.** Fits of the resistivity data for La$_{0.7-x}$Ln$_x$Ca$_{0.3}$CoO$_3$ for x > x$_c$ to the T$^{-1/4}$ law in the 49 – 200 K range. The symbols represent experimental data points and broken lines represent the linear fits.

**Table 1. Crystal Structure data of $La_{0.7-x}Ln_xCa_{0.3}CoO_3$ (Ln = Pr, Nd, Gd or Dy)**

| Composition | $\langle r_A \rangle$ (Å) | Space Group | Lattice parameters (Å) a | b | c | V ($Å^3$) |
|---|---|---|---|---|---|---|
| x = 0.0 | 1.354 | R $\bar{3}$C | 5.3906 | - | - | 111.6 |
| Ln = Pr | | | | | | |
| x = 0.1 | 1.347 | R $\bar{3}$C | 5.3869 | - | - | 111.2 |
| x = 0.2 | 1.340 | R $\bar{3}$C | 5.3837 | - | - | 111.0 |
| x = 0.3 | 1.194 | Pnma | 5.3836 | 7.5858 | 5.3679 | 219.2 |
| x = 0.5 | 1.187 | Pnma | 5.3723 | 7.5686 | 5.3663 | 218.2 |
| x = 0.6 | 1.183 | Pnma | 5.3652 | 7.5731 | 5.3593 | 217.8 |
| x = 0.7 | 1.179 | Pnma | 5.3577 | 7.5774 | 5.3436 | 216.9 |
| Ln = Nd | | | | | | |
| x = 0.1 | 1.345 | R $\bar{3}$C | 5.3784 | - | - | 110.7 |
| x = 0.2 | 1.336 | R $\bar{3}$C | 5.3761 | - | - | 110.6 |
| x = 0.3 | 1.189 | Pnma | 5.3795 | 7.5867 | 5.3732 | 219.3 |
| x = 0.4 | 1.184 | Pnma | 5.3700 | 7.5741 | 5.3642 | 218.4 |
| x = 0.49 | 1.179 | Pnma | 5.3667 | 7.5766 | 5.3641 | 218.1 |
| x = 0.7 | 1.168 | Pnma | 5.346 | 7.5638 | 5.3287 | 215.5 |

**Table 1 Continued**

| Ln = Gd | | | | | | |
|---|---|---|---|---|---|---|
| x = 0.1 | 1.338 | R $\bar{3}$C | 5.3797 | - | - | 110.7 |
| x = 0.2 | 1.322 | R $\bar{3}$C | 5.3785 | - | - | 110.4 |
| x = 0.24 | 1.179 | Pnma | 5.3838 | 7.5846 | 5.3797 | 219.7 |
| x = 0.3 | 1.1725 | Pnma | 5.3813 | 7.5814 | 5.3624 | 218.8 |
| Ln = Dy | | | | | | |
| x = 0.1 | 1.337 | R $\bar{3}$C | 5.3759 | - | - | 110.1 |
| x = 0.195 | 1.179 | Pnma | 5.3960 | 7.5990 | 5.3729 | 220.3 |
| x = 0.3 | 1.165 | Pnma | 5.3813 | 7.5783 | 5.3583 | 218.5 |

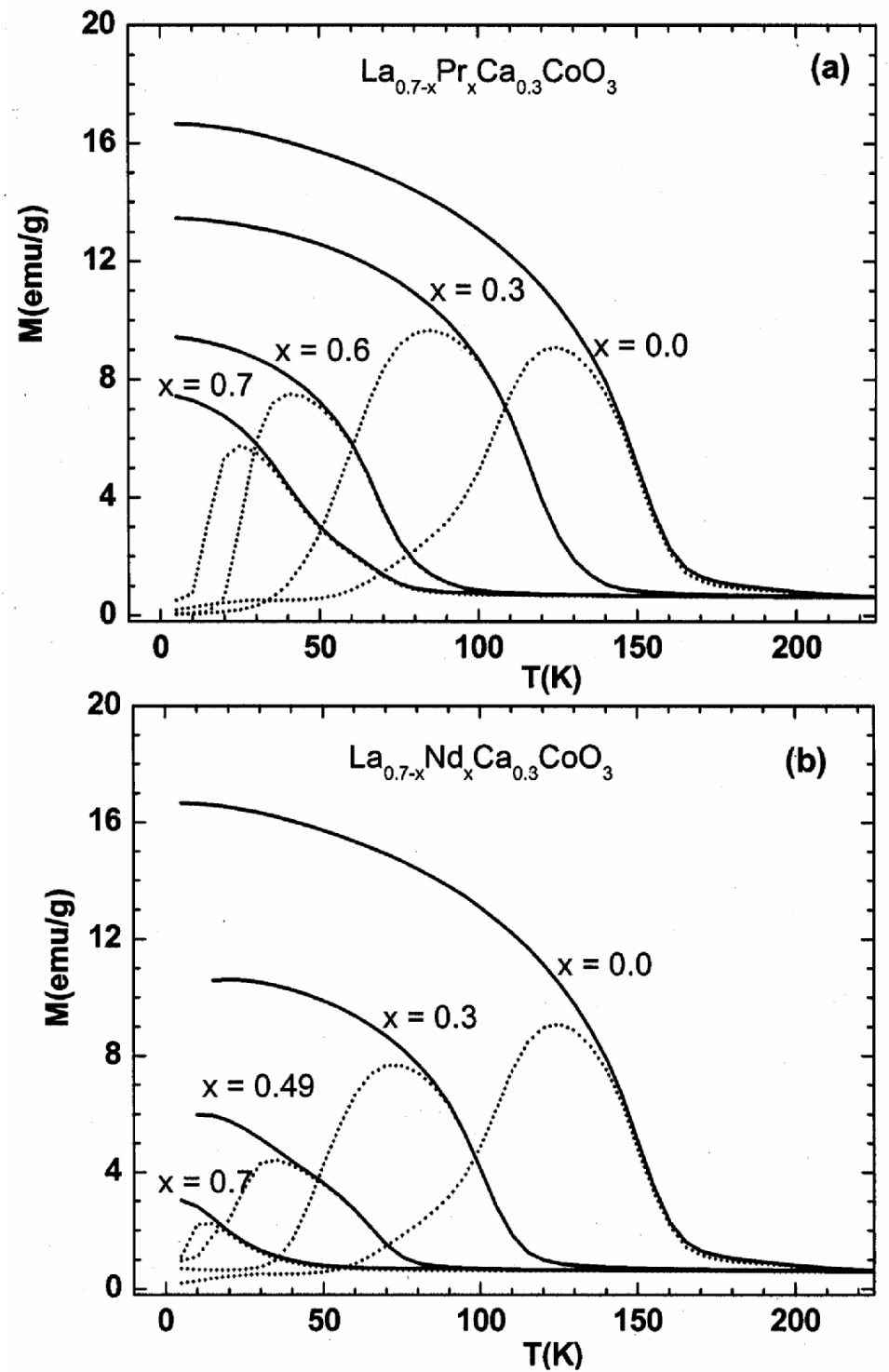

**Figure 1**

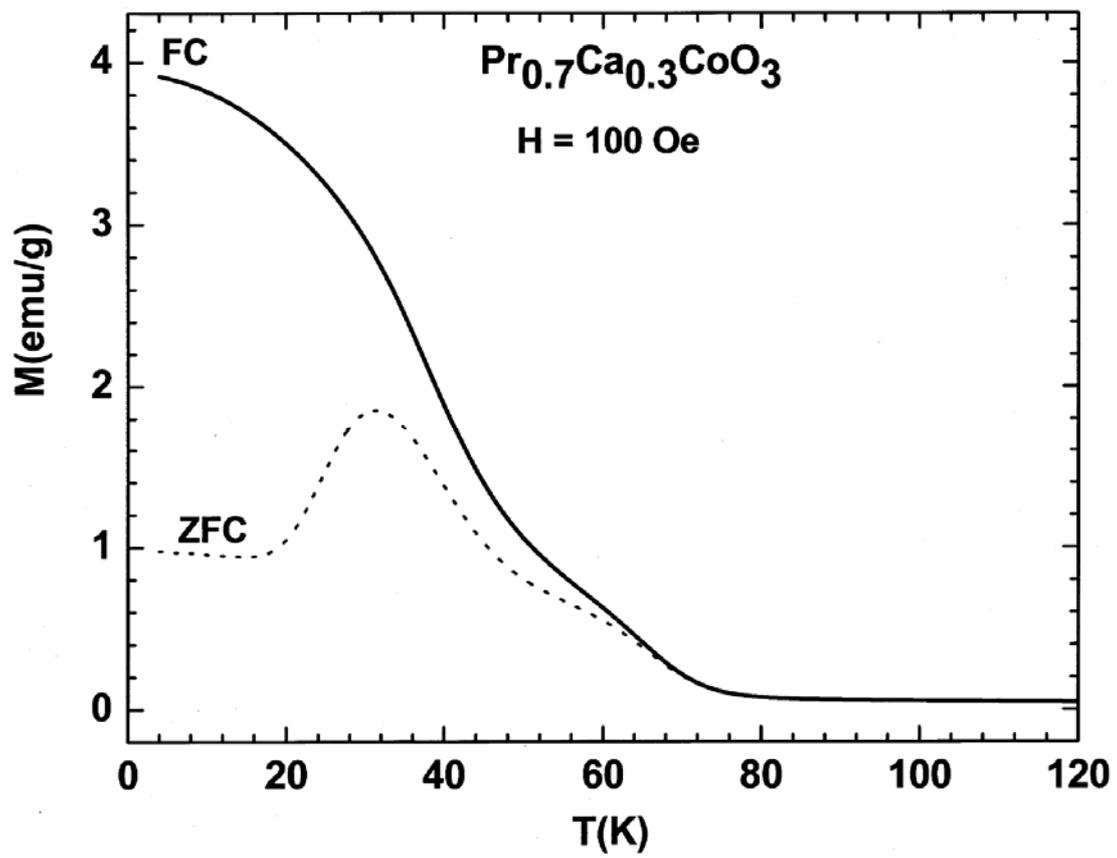

**Figure 2**

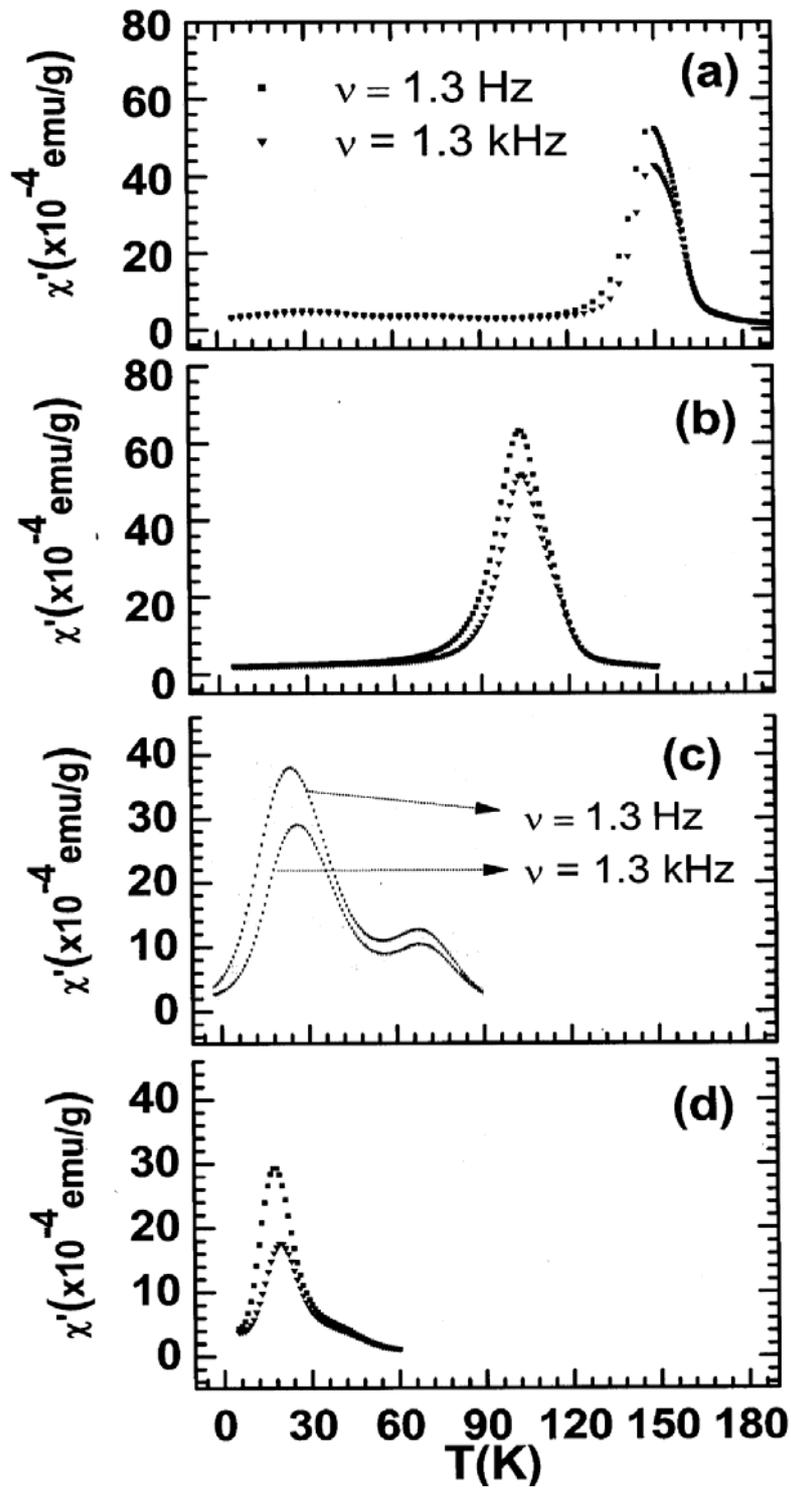

**Figure 3**

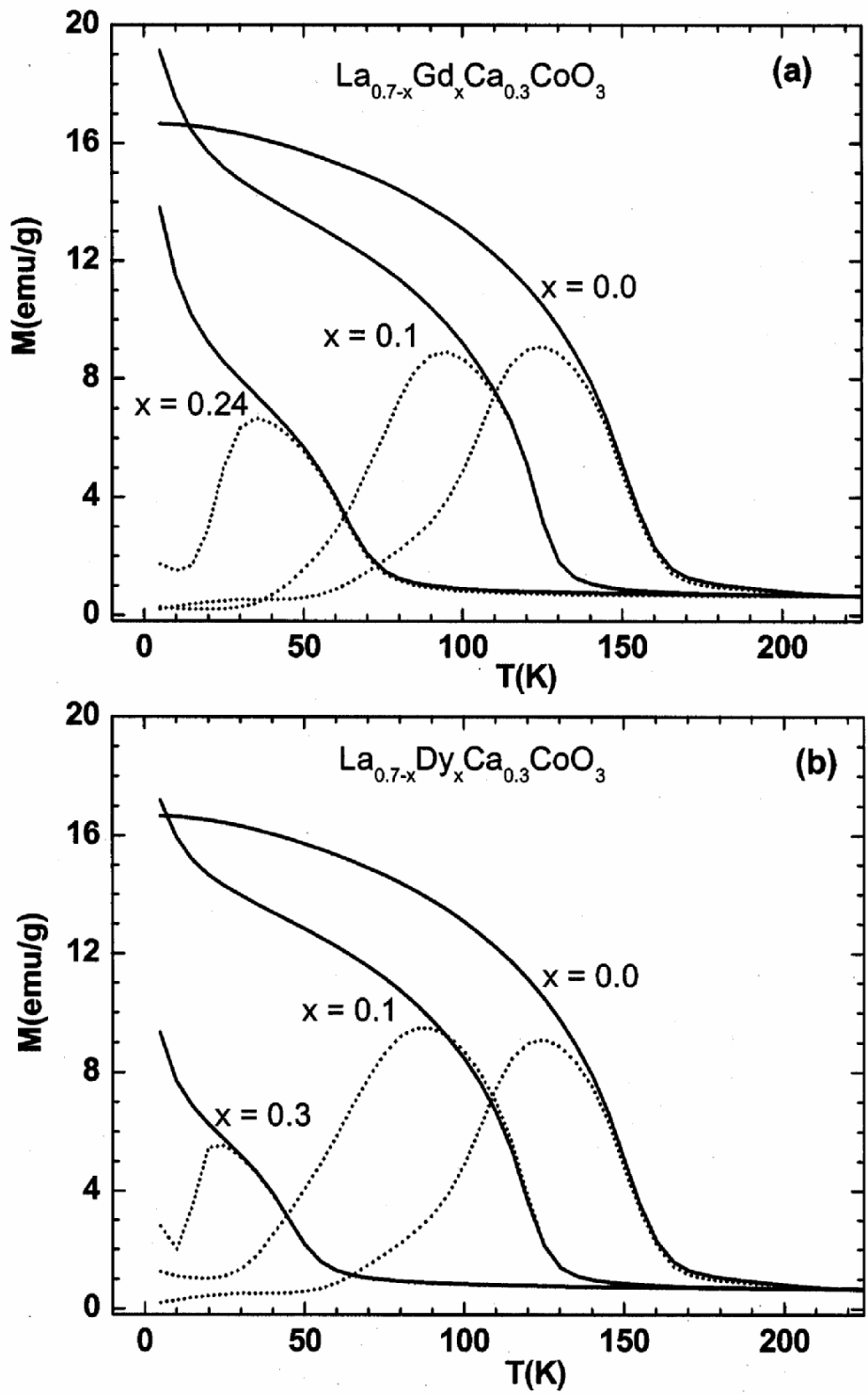

**Figure 4**

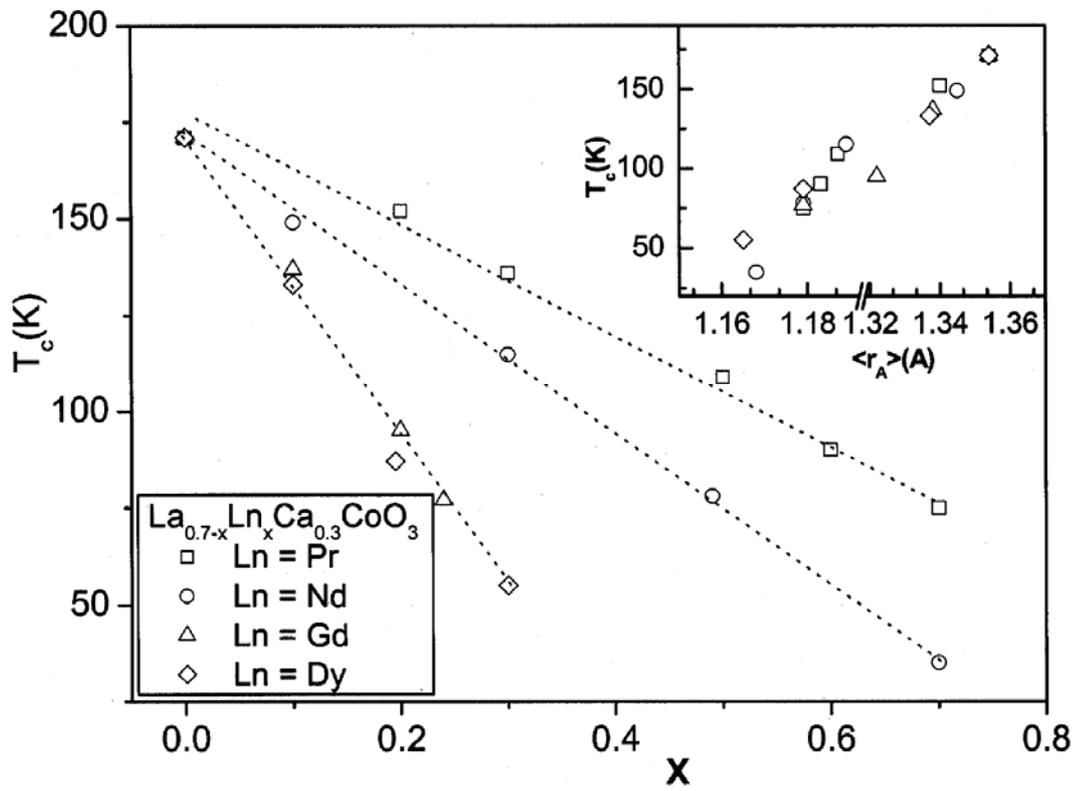

**Figure 5**

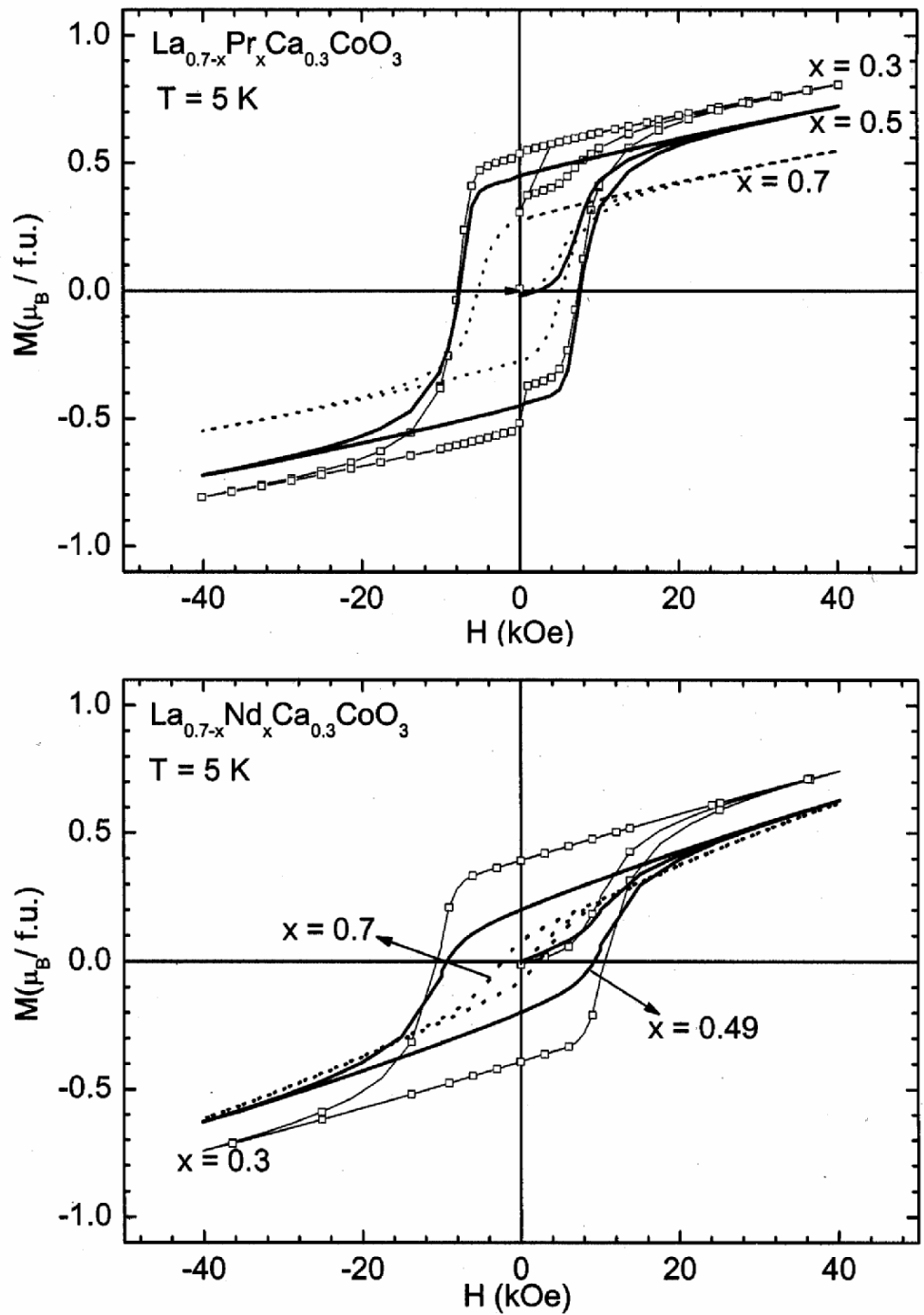

**Figure 6**

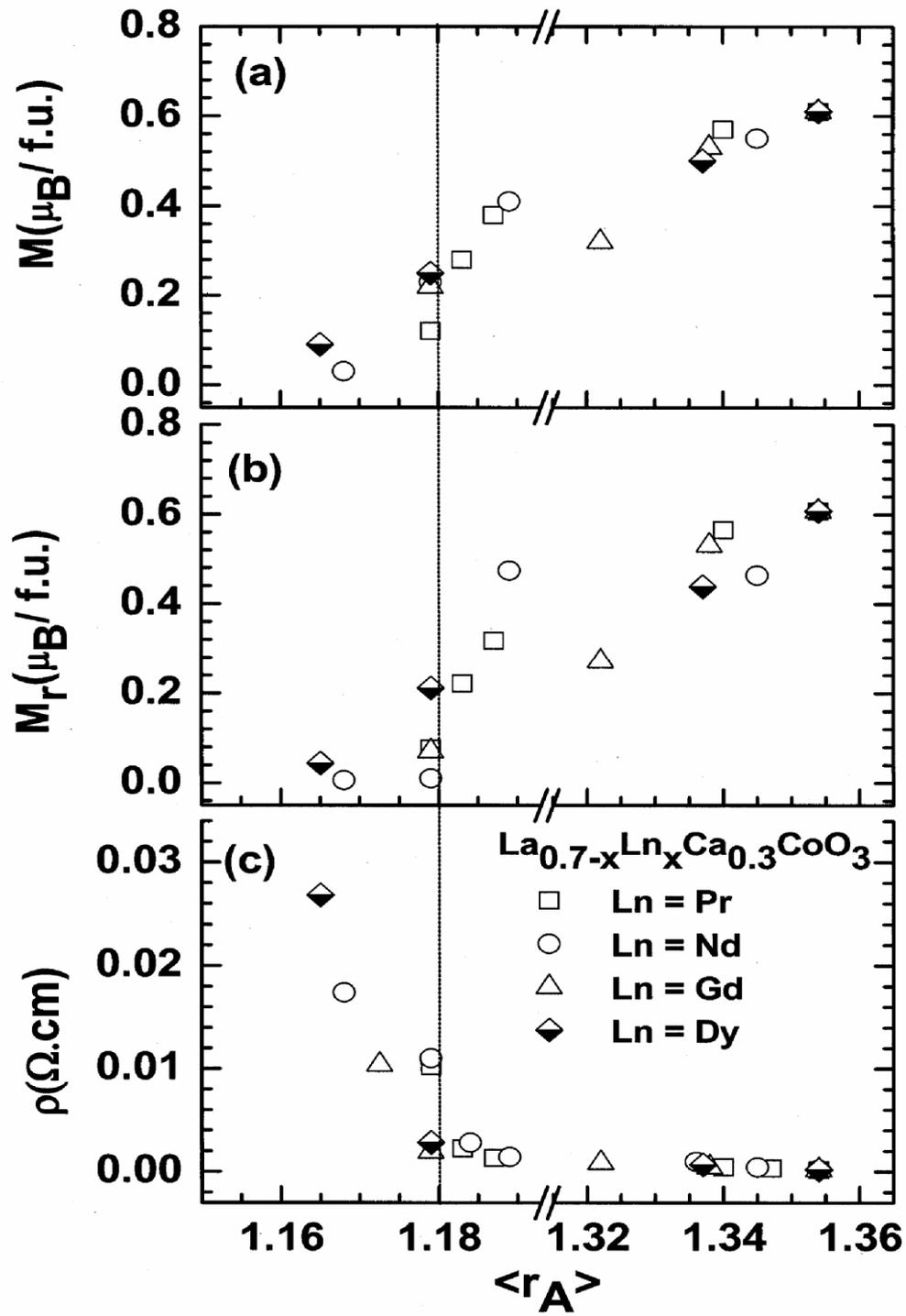

**Figure 7**

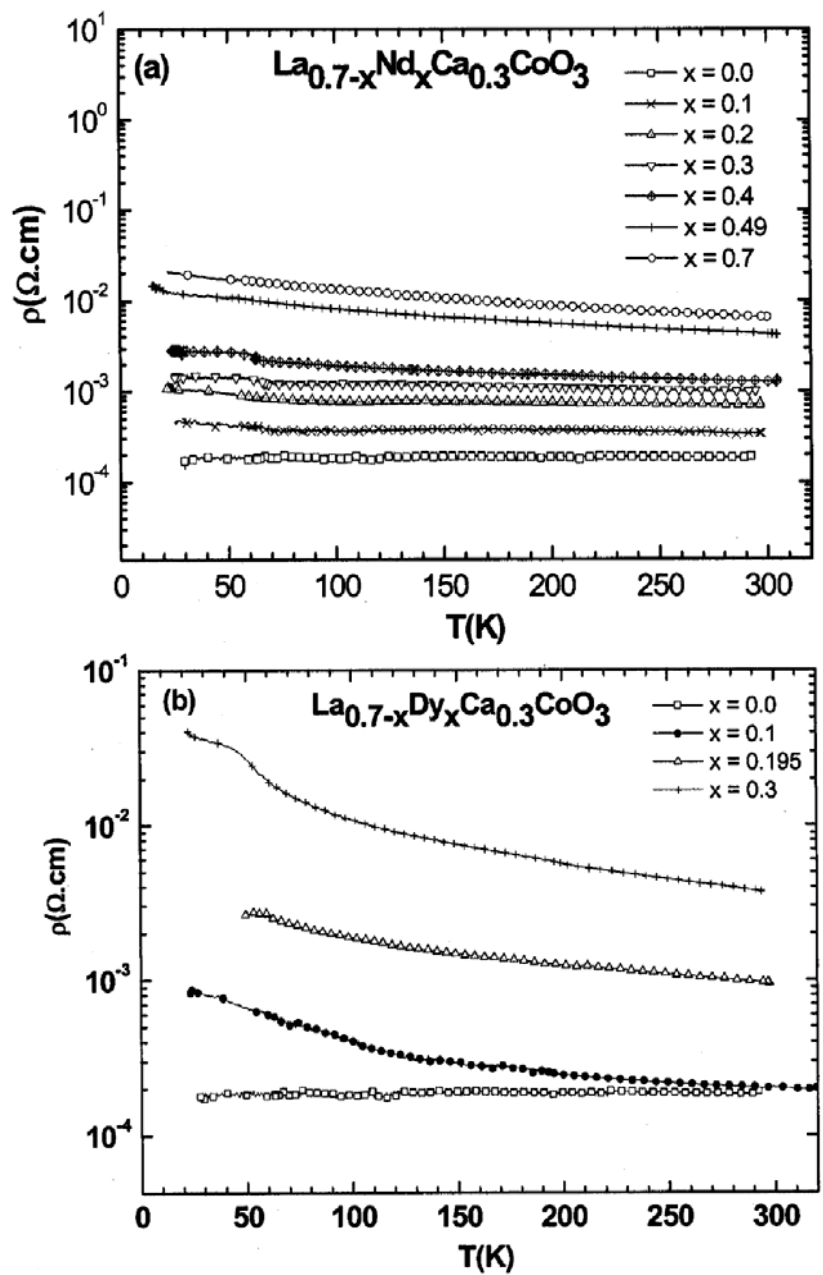

**Figure 8**

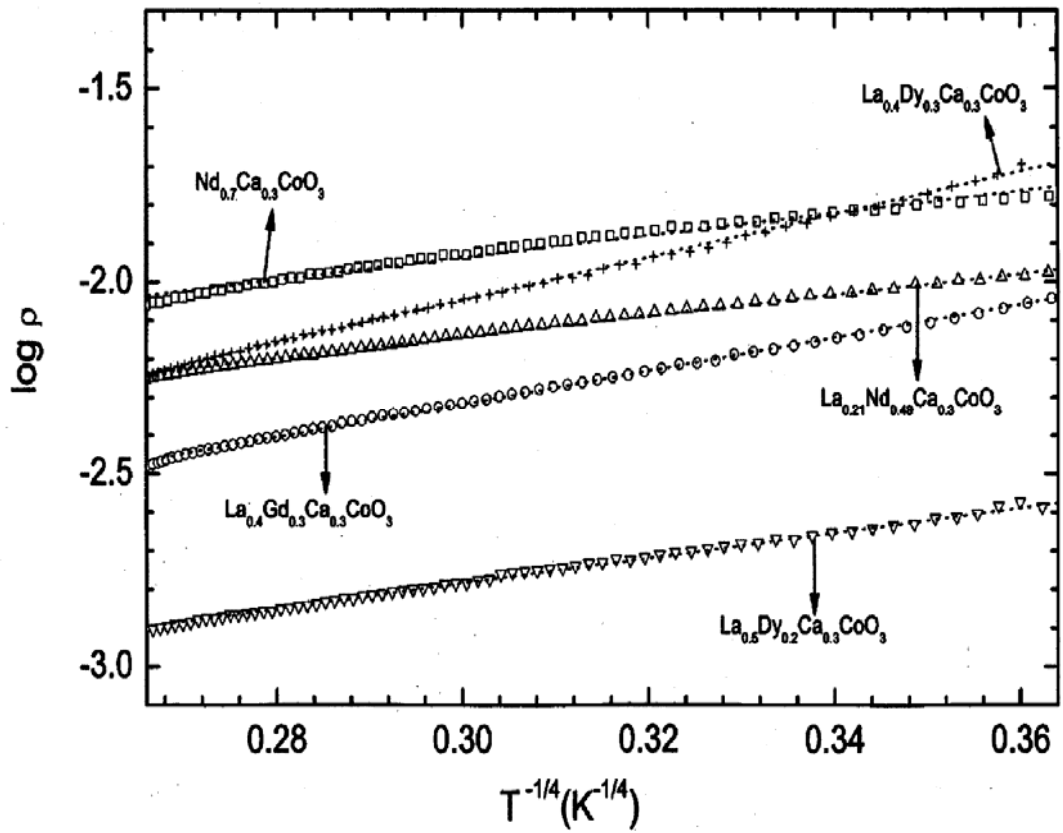

Figure 9